\documentclass[11pt]{article}
\usepackage[T1]{fontenc}
\usepackage{textcomp}
\usepackage[latin9]{inputenc}
\usepackage{amsmath}
\usepackage{amssymb}
\usepackage[a4paper]{geometry}
\usepackage{esint}

\makeatletter

\pdfoutput=1 

\usepackage{jheppub}

\usepackage{etoolbox}
    
    \patchcmd{\maketitle}{\@fpheader}{}{}{}

\usepackage{amsfonts}

\usepackage{bbm}

\title{\boldmath Enhanced Conformal BMS$_{3}$ Symmetries }

\author[a,b,c]{Oscar Fuentealba,}
\author[d]{Iva Lovrekovic,}
\author[a,b]{David Tempo,}
\author[e,f]{Ricardo Troncoso}

\affiliation[a]{Instituto de Ciencias Exactas y Naturales (ICEN), Universidad Arturo Prat, Playa Brava 3256, 1111346 Iquique, Chile}
\affiliation[b]{Facultad de Ciencias, Universidad Arturo Prat, Avenida Arturo Prat Chac\'on 2120, 1110939 Iquique, Chile}
\affiliation[c]{International Solvay Institutes, ULB-Campus Plaine CP231, B-1050 Brussels, Belgium}
\affiliation[d]{Institute for Theoretical Physics, TU Wien, Wiedner Hauptstrasse, 8-10, 1040, Vienna, Austria}
\affiliation[e]{Facultad de Ingenier\'{i}a, Arquitectura y Dise\~{n}o, Universidad San Sebasti\'{a}n, sede Valdivia,
General Lagos 1163, Valdivia 5110693, Chile}
\affiliation[f]{Centro de Estudios Cient\'{i}ficos (CECs), Av. Arturo Prat 514, Valdivia, Chile}

\emailAdd{ofuentealba@unap.cl}
\emailAdd{iva.lovrekovic@tuwien.ac.at}
\emailAdd{jtempo@unap.cl}
\emailAdd{ricardo.troncoso@uss.cl}

\preprint{TUW 24-06}

\abstract{An enhanced version of the conformal BMS$_{3}$ algebra is presented.
It is shown to emerge from the asymptotic structure of an extension
of conformal gravity in 3D by Pope and Townsend that consistently
accommodates an additional spin-2 field, once it is endowed
with a suitable set of boundary conditions. The canonical generators
of the asymptotic symmetries then span a precise nonlinear
W$_{(2,2,2,2,1,1,1)}$ algebra, whose central extensions and coefficients
of the nonlinear terms are completely determined by the central charge
of the Virasoro subalgebra. The wedge algebra corresponds to the conformal
group in four dimensions $SO(4,2)$ and therefore, enhanced conformal
BMS$_{3}$  can also be regarded as an infinite-dimensional nonlinear
extension of the AdS$_{5}$ algebra with nontrivial central extensions.
It is worth mentioning that our boundary conditions might be considered
as a starting point in order to consistently incorporate either a finite
or an infinite number of conformal higher spin fields.}

\makeatother

\begin{document}
\maketitle \flushbottom

\section{Introduction}

Finding a bona fide conformal extension of the BMS algebra appears
to be a hard nut to crack (see e.g. \cite{Haco:2017ekf,Adami:2020ugu,Donnay:2020fof,Batlle:2020hia,Bekaert:2022oeh}).
Nevertheless in three spacetime dimensions the task can be successfully
achieved, provided that an infinite number of superdilatations and
superspecial conformal transformations are incorporated within a nonlinear
algebra \cite{Fuentealba:2020zkf}. Specifically, the commutator of
supertranslations with superspecial conformal transformations acquires
quadratic and cubic terms made of superrotations and superdilatations.
The conformal BMS$_{3}$ algebra has been shown to emerge in different
physical setups, as it is the case of the asymptotic symmetry algebra
of conformal gravity in 3D \cite{Fuentealba:2020zkf}, as well as
from the free field realization of the BMS$_{3}$ Ising model in 2D
\cite{Yu:2022bcp}. Further related results can be found in \cite{Gupta:2023fmp,Gupta:2024gcn}.

The conformal BMS$_{3}$ algebra seems to be very rigid, since the
central extensions and the coefficients of the nonlinear terms become
entirely determined by the central charge of the Virasoro subalgebra.
Indeed, the conditions obtained from the Jacobi identity turn out
to be very stringent, which suggests that the algebra is unique. In
this sense, since the conformal BMS$_{3}$ algebra looks undeformable,
one may wonder whether it might be enhanced in some appropriate way.
As a strategy to achieve this task we propose exploring the asymptotic
structure of a suitable extension of conformal gravity in 3D \cite{vanNieuwenhuizen:1985cx,Horne:1988jf}.
A nice and simple theory enjoying the sought features was proposed
long ago by Pope and Townsend \cite{Pope:1989vj} with the aim of
further enlarging it in order to describe an infinite tower of conformal
higher spin fields in 3D, along the lines of \cite{Fradkin:1987ks}.
More recently, conformal gravity was shown to admit a different extension
that accommodates a large class of theories with a finite number of conformal higher spin fields
\cite{Grigoriev:2019xmp}.\footnote{Additional interesting results concerning conformal higher spin fields in 3D
can be found in e.g., \cite{Nilsson:2013tva,Henneaux:2015cda,Basile:2017mqc,Kuzenko:2020opc,Diaz:2024iuz,Lovrekovic:2024yoo,Grigoriev:2020lzu}.}

The theory proposed in \cite{Pope:1989vj} describes a non-gauged
spin-2 field consistently coupled to conformal gravity, and it can
also be formulated in terms of a Chern-Simons action for $so(4,2)$,
after a suitable gauge choice akin to that of Horne and Witten for
the case of pure conformal gravity \cite{Horne:1988jf}.

In the next section we show that the searched-for enhancement of the
conformal BMS$_{3}$ algebra naturally emerges from the asymptotic
structure of the extension of conformal gravity aforementioned.

\section{Asymptotic structure of extended conformal gravity in 3D}

Following Pope and Townsend \cite{Pope:1989vj}  the three-dimensional
conformal algebra $so(3,2)$, spanned by $\left\{ J_{a},P_{a},K_{a},D\right\} $
with $a=0,1,2$, is enhanced to that of $so(4,2)$ by enlarging the
set of generators to include $\left\{ U_{a},U,V\right\} $. This is also isomorphic to the algebra of depth-two conformal gravity in Grigoriev et al. \cite{Grigoriev:2019xmp}.

For our purposes, it is convenient to arrange the generators in a
different way. The generators of the $so(2,1)\approx sl(2,\mathbb{R})$
subalgebra that commutes with the Lorentz subalgebra (spanned by $J_{a}$)
are given by $T^{I}=\left\{ U,V,D\right\} $ with $I=0,1,2$, so that
the remaining generators $P_{a}^{I}=\left\{ P_{a},K_{a},U_{a}\right\} $
also behave like vectors of $sl(2,\mathbb{R})$. The full $so(4,2)$
algebra then explicitly reads
\begin{align}
\left[J_{a},J_{b}\right] & =\epsilon_{abc}J^{c}\,\,\,;\,\,\,\left[J_{a},P_{b}^{I}\right]=\epsilon_{ab}^{\;\;\;\;c}P_{c}^{I}\;,\nonumber \\
\left[T^{I},T^{J}\right] & =\epsilon^{IJK}T_{K}\,\,\,;\,\,\,\left[T^{I},P_{a}^{J}\right]=\epsilon_{\;\;\;\;K}^{IJ}P_{a}^{K}\;,\label{eq:so(4,2)algebra}\\
\left[J_{a},T^{I}\right] & =0\,\,\,;\,\,\,\left[P_{a}^{I},P_{b}^{J}\right]=-2\tau^{IJ}\epsilon_{abc}J^{c}-2\eta_{ab}\epsilon^{IJK}T_{K}\;,\nonumber 
\end{align}
where both $\eta_{ab}$ and $\tau_{IJ}$ stand for the flat Minkowski
metric in 3D. It is useful to express $\tau^{IJ}$ in light cone coordinates
($\tau^{01}=\tau^{10}=\tau^{22}=1$) so that Poincaré translations
and special conformal transformations correspond to $P_{a}=P_{a}^{0}$
and $K_{a}=P_{a}^{1}$, respectively; while dilations do for $D=T^{2}$. 

We choose the normalization of the Cartan-Killing metric so that it
reads
\begin{equation}
\left\langle J_{a}J_{b}\right\rangle =\eta_{ab}\,\,\,;\,\,\,\left\langle P_{a}^{I}P_{b}^{J}\right\rangle =-2\eta_{ab}\tau^{IJ}\,\,\,;\,\,\,\left\langle T_{I}T_{J}\right\rangle =\tau_{IJ}\,,\label{eq:CartanKilling}
\end{equation}
and hence, the extension of conformal gravity in 3D can be expressed
in terms of a Chern-Simons action 
\begin{equation}
I\left[A\right]=\frac{k}{4\pi}\int\left\langle AdA+\frac{2}{3}A^{3}\right\rangle \thinspace,
\end{equation}
for a gauge field given by
\begin{equation}
A=\omega^{a}J_{a}+E_{I}^{a}P_{a}^{I}+M_{I}T^{I}\;,
\end{equation}
where $e^{a}=E_{0}^{a}$ and $\omega^{a}$ stand for the dreibein
and the dualized spin connection, while $s^{a}=E_{2}^{a}$ corresponds
to the one-form associated to the spin-2 gauge field \cite{Grigoriev:2019xmp}. \footnote{This field was initially identified as a spin-2 non-gauge field \cite{Pope:1989vj}.}

\subsection{Boundary conditions and enhanced conformal BMS$_{3}$ algebra}

Following the lines of \cite{Coussaert:1995zp}, a gauge choice of
the form $A=g^{-1}ag+g^{-1}dg$, with a suitable group element $g=g\left(r\right)$,
allows to completely gauge away the radial dependence of the asymptotic
form of the connection, so that the remaining analysis can be readily
performed in terms of the auxiliary gauge field $a=a_{t}dt+a_{\varphi}d\varphi$
that depends only on time and the angular coordinate.

We then propose the following fall-off of the gauge field 
\begin{align}
a & =\left[J_{1}+\frac{2\pi}{k}\left(\mathcal{J}+\frac{2\pi}{k}\Lambda_{(2)}\right)J_{0}+\frac{\pi}{k}\mathcal{P}_{I}P_{0}^{I}+\frac{2\pi}{k}\mathcal{M}_{I}T^{I}\right]\left(d\varphi+dt\right)\,,\label{eq:a}
\end{align}
where $\Lambda_{(2)}=\tau_{IJ}\Lambda_{(2)}^{IJ}$, with
\begin{equation}
\Lambda_{(2)}^{IJ}=\frac{1}{4}\left({\cal M}^{I}{\cal M}^{J}-\tau^{IJ}{\cal M}^{K}{\cal M}_{K}\right)\;,
\end{equation}
and the dynamical fields $\mathcal{J}$, $\mathcal{P}_{I}$, $\mathcal{M}_{J}$
depend only on $t$, $\varphi$.

The asymptotic behavior is preserved under gauge transformations $\delta a=d\Omega+\left[a,\Omega\right]$,
with a Lie-algebra-valued parameter given by
\begin{align}
\Omega\left[\epsilon,\zeta^{I},\lambda^{J}\right] & =\epsilon J_{1}-\zeta_{I}P_{1}^{I}-\left(\lambda_{I}-\frac{2\pi}{k}{\cal M}_{I}\epsilon\right)T^{I}+\Theta\left[\epsilon,\zeta^{I},\lambda^{J}\right]\;,
\end{align}
with 
\begin{align}
\Theta= & -\epsilon^{\prime}J_{2}+\frac{2\pi}{k}\left(\zeta_{K}^{\prime}-\frac{2\pi}{k}\epsilon_{\;\;\;\;K}^{IJ}\zeta_{I}\mathcal{M}_{J}\right)P_{2}^{K}+\frac{2\pi}{k}\left[\epsilon\left(\mathcal{J}+\frac{2\pi}{k}\Lambda_{(2)}\right)+\zeta_{I}{\cal P}^{I}-\frac{k}{2\pi}\epsilon^{\prime\prime}\right]J_{0}\nonumber \\
 & -\frac{2\pi}{k}\left[\zeta_{K}\left(\mathcal{J}+\frac{2\pi}{k}\Lambda_{(2)}\right)+\frac{8\pi}{k}\zeta_{I}\Lambda_{(2)K}^{I}+\epsilon_{\;\;\;\;K}^{IJ}\left(\zeta_{I}\mathcal{M}_{J}^{\prime}+2\zeta_{I}^{\prime}\mathcal{M}_{J}\right)-\frac{1}{2}\epsilon{\cal P}_{K}-\frac{k}{2\pi}\zeta_{K}^{\prime\prime}\right]P_{0}^{K}\,,
\end{align}
depending on chiral functions of $t$, $\varphi$ fulfilling $\dot{\epsilon}=\epsilon^{\prime}$,
$\dot{\zeta}_{I}=\zeta_{I}^{\prime}$, $\dot{\lambda}_{I}=\lambda_{I}^{\prime}$.
Note that anti-chiral functions would be obtained if the asymptotic
behavior of the gauge field in \eqref{eq:a} were chosen as a one-form
along $d\varphi-dt$ (instead of $d\varphi+dt$). 

The transformation law of the dynamical fields is then given by
\begin{align}
\delta{\cal J} & =2\mathcal{J}\epsilon^{\prime}+\mathcal{J}^{\prime}\epsilon-\frac{k}{2\pi}\epsilon^{\prime\prime\prime}+2\mathcal{P}^{I}\zeta_{I}^{\prime}+2\mathcal{P}_{I}^{\prime}\zeta^{I}-\mathcal{M}^{I}\lambda_{I}^{\prime}\;,\nonumber \\
\delta\mathcal{P}^{I} & =2\mathcal{P}^{I}\epsilon^{\prime}+\mathcal{P}^{I\prime}\epsilon-4\left(\mathcal{J}\tau^{IJ}-\tilde{\Lambda}_{(2)}^{IJ}\right)\zeta_{J}^{\prime}-2\left[\left(\mathcal{J}\tau^{IJ}-\tilde{\Lambda}_{(2)}^{IJ}\right)^{\prime}+\epsilon^{IJK}{\cal M}_{K}^{\prime\prime}+\frac{1}{2}\tilde{\Lambda}_{(3)}^{IJ}\right]\zeta_{J}\nonumber \\
 & \quad+6\left(\epsilon^{IJK}{\cal M}_{J}\zeta_{K}^{\prime}\right)^{\prime}-\epsilon^{IJK}{\cal P}_{J}\lambda_{K}+\frac{k}{\pi}\zeta^{I\prime\prime\prime}\;,\label{eq:TransLaws}\\
\delta\mathcal{{\cal M}}_{I} & =\mathcal{{\cal M}}_{I}\epsilon^{\prime}+\mathcal{{\cal M}}_{I}^{\prime}\epsilon+\epsilon_{IJK}{\cal P}^{J}\zeta^{K}-\epsilon_{I}^{\;\;JK}{\cal M}_{J}\lambda_{K}-\frac{k}{2\pi}\lambda_{I}^{\prime}\thinspace.\nonumber 
\end{align}
with $\tilde{\Lambda}_{(2)}^{IJ}$ and $\tilde{\Lambda}_{(3)}^{IJ}$
being symmetric and antisymmetric in $I$, $J$, respectively, and
defined as
\begin{align}
\tilde{\Lambda}_{(2)}^{IJ} & =-\frac{2\pi}{k}\left(\Lambda_{(2)}\tau^{IJ}+6\Lambda_{(2)}^{IJ}\right)\;,\nonumber \\
\tilde{\Lambda}_{(3)}^{IJ} & =-\frac{4\pi}{k}\left[2\left({\cal J}+\frac{4\pi}{k}\Lambda_{(2)}\right)\epsilon^{IJK}{\cal M}_{K}+{\cal M}^{\left[I\right.}{\cal M}^{\prime\left.J\right]}\right]\;.\label{eq:Lambda2,3}
\end{align}
The generators of the asymptotic symmetries can then be straightforwardly
obtained following different approaches, as in \cite{Regge:1974zd,Barnich:2001jy} (see also e.g., \cite{Balachandran:1991dw,Banados:1994tn,Carlip:2005zn}), and they are given by
\begin{align}
\mathcal{Q}\left[\epsilon,\zeta_{I},\lambda_{I}\right] & =-\int\left(\epsilon\mathcal{J}+\zeta_{I}\mathcal{P}^{I}-\lambda_{I}{\cal M}^{I}\right)d\varphi\,,\label{eq:QExt}
\end{align}
so that their algebra can be extracted from their Dirac brackets;
or in a more direct way, by virtue of $\delta_{\eta_{1}}{\cal Q}\left[\eta_{2}\right]=\left\{ {\cal Q}\left[\eta_{2}\right],{\cal Q}\left[\eta_{1}\right]\right\} $,
and the transformation law of the dynamical fields in \eqref{eq:TransLaws}. 

The algebra of the asymptotic symmetry generators is then found to
be described by 
\begin{align}
\left\{ {\cal J}(\phi),{\cal J}(\varphi)\right\}  & =-2\mathcal{J}(\phi)\delta^{\prime}\left(\phi-\varphi\right)-\delta\left(\phi-\varphi\right)\mathcal{J}^{\prime}(\phi)+\frac{k}{2\pi}\delta^{\prime\prime\prime}\left(\phi-\varphi\right)\;,\nonumber \\
\left\{ {\cal J}(\phi),{\cal P}^{K}(\varphi)\right\}  & =-2\mathcal{P}^{K}(\phi)\delta^{\prime}\left(\phi-\varphi\right)-\delta\left(\phi-\varphi\right)\mathcal{P}^{K\prime}(\phi)\;,\nonumber \\
\left\{ {\cal J}(\phi),{\cal M}^{I}(\varphi)\right\}  & =-{\cal M}^{I}(\phi)\delta^{\prime}\left(\phi-\varphi\right)\;,\nonumber \\
\left\{ {\cal M}^{I}(\phi),{\cal M}^{J}(\varphi)\right\}  & =\epsilon^{IJK}{\cal M}_{K}\delta\left(\phi-\varphi\right)-\frac{k}{2\pi}\tau^{IJ}\delta^{\prime}\left(\phi-\varphi\right)\;,\label{eq:AlgebraDeltas}\\
\left\{ {\cal P}_{I}(\phi),{\cal M}_{J}(\varphi)\right\}  & =\epsilon_{IJK}{\cal P}^{K}(\phi)\delta\left(\phi-\varphi\right)\;,\nonumber \\
\left\{ {\cal P}^{I}(\phi),{\cal P}^{J}(\varphi)\right\}  & =4\left(\mathcal{J}(\phi)\tau^{IJ}-\tilde{\Lambda}_{(2)}^{IJ}(\phi)\right)\delta^{\prime}\left(\phi-\varphi\right)\nonumber \\
 & \quad+2\left(\mathcal{J}(\phi)\tau^{IJ}-\tilde{\Lambda}_{(2)}^{IJ}(\phi)+\epsilon^{IJK}\mathcal{M}_{K}^{\prime}(\phi)\right)^{\prime}\delta\left(\phi-\varphi\right)\nonumber \\
 & \quad-6\left(\epsilon^{IJK}\mathcal{M}_{K}(\phi)\delta^{\prime}\left(\phi-\varphi\right)\right)^{\prime}+\tilde{\Lambda}_{(3)}^{IJ}(\phi)\delta\left(\phi-\varphi\right)-\frac{k}{\pi}\tau^{IJ}\delta^{\prime\prime\prime}\left(\phi-\varphi\right)\;.\nonumber 
\end{align}
Expanding in Fourier modes according to $X=\frac{1}{2\pi}\sum_{m}X_{m}e^{im\varphi}$,
the algebra reads
\begin{align}
i\left\{ \mathcal{J}_{m},\mathcal{J}_{n}\right\}  & =\left(m-n\right)\mathcal{J}_{m+n}+m\left(m^{2}-1\right)k\delta_{m+n,0}\,,\nonumber \\
i\left\{ \mathcal{J}_{m},\mathcal{P}_{n}^{I}\right\}  & =\left(m-n\right)\mathcal{P}_{m+n}^{I}\,,\nonumber \\
i\left\{ \mathcal{J}_{m},\mathcal{M}_{n}^{I}\right\}  & =-n\mathcal{M}_{m+n}^{I}\,,\nonumber \\
i\left\{ {\cal M}_{m}^{I},{\cal M}_{n}^{J}\right\}  & =i\epsilon_{\;\;\;\;\;K}^{IJ}{\cal M}_{m+n}^{K}+k\tau^{IJ}m\delta_{m+n,0}\;,\label{eq:AlgebraModes}\\
i\left\{ \mathcal{P}_{m}^{I},\mathcal{M}_{n}^{J}\right\}  & =i\epsilon_{\;\;\;\;K}^{IJ}\mathcal{P}_{m+n}^{K}\,,\nonumber \\
i\left\{ \mathcal{P}_{m}^{I},\mathcal{P}_{n}^{J}\right\}  & =-2\left(m-n\right)\mathcal{J}_{m+n}\tau^{IJ}+\left(m-n\right)\tilde{\Lambda}_{(2)m+n}^{IJ}\nonumber \\
 & \quad-2i\left(m^{2}-mn+n^{2}-1\right)\epsilon_{\;\;\;\;Q}^{IJ}\mathcal{M}_{m+n}^{Q}+\tilde{\Lambda}_{(3)m+n}^{IJ}+2k\tau^{IJ}m\left(m^{2}-1\right)\delta_{m+n,0}\,,\nonumber 
\end{align}
where the zero mode of ${\cal J}_{n}$ has been shifted as ${\cal J}_{0}\rightarrow{\cal J}_{0}-\frac{k}{4\pi}$,
and the nonlinear terms given by
\begin{align}
\tilde{\Lambda}_{(2)m}^{IJ} & =\frac{4}{k}\tau^{IJ}\tau_{KL}\sum_{n}{\cal M}_{m-n}^{K}{\cal M}_{n}^{L}-\frac{3}{k}\sum_{n}{\cal M}_{m-n}^{I}{\cal M}_{n}^{J}\;,\nonumber \\
\tilde{\Lambda}_{(3)m}^{IJ} & =-\frac{4}{k}\epsilon_{\;\;\;K}^{IJ}\sum_{p}\left({\cal J}_{m+p}-\frac{1}{k}\tau_{QR}\sum_{n}{\cal M}_{m-n-p}^{Q}{\cal M}_{n}^{R}\right){\cal M}_{p}^{K}+\frac{i}{2k}\sum_{n}n{\cal M}_{n}^{[I}{\cal M}_{m-n}^{J]}\;,\label{eq:Lambda2,3modes}
\end{align}
possess (anomalous) conformal weight $2$ and $3$, respectively.
Indeed, the conformal weight of the generators ${\cal J}_{m}$, ${\cal P}_{m}^{I}$
is 2, while that of the currents ${\cal M}_{m}^{I}$ is clearly 1.

The wedge algebra is then given by that of the original gauge group
$SO(4,2)$ in \eqref{eq:so(4,2)algebra}, being recovered once the
nonlinear terms are dropped and the modes are restricted according
to $\left|m\right|<s$, where $s$ stands for the conformal weight
of the generators, followed by a simple change of basis. 

Note that since the Lorentz subalgebra is non-principally embedded
within the wedge algebra, according to the conformal weight of the
generators, the enhanced conformal BMS$_{3}$ algebra \eqref{eq:AlgebraModes}
can be regarded as a $W_{(2,2,2,2,1,1,1)}$ algebra (see e.g. \cite{Bouwknegt:1995ag,Frappat:1992bs}). 

It is also worth highlighting that the Virasoro central charge gives
support to the nonlinear terms, and therefore, the enhanced conformal
BMS$_{3}$ algebra turns out to be well-defined provided the central
charge does not vanish. Nonetheless, for the quantum algebra this
is not necessarily the case because the central extensions and the
coefficient in front of the nonlinear terms generically acquire corrections.

\section{Ending remarks}

The enhanced conformal BMS$_{3}$ algebra \eqref{eq:AlgebraModes}
inherits the ``rigidity'' of its non enhanced version in \cite{Fuentealba:2020zkf},
since all of the central extensions and the coefficients of the nonlinear
terms also become entirely fixed in terms of the central charge of
the Virasoro subalgebra, determined by the Chern-Simons level $k$.
This can be traced back by the fact that the extension of the conformal
algebra $so(4,2)$ is semisimple, so that it possesses a unique invariant
bilinear form given by the Cartan-Killing metric that can be normalized
as in \eqref{eq:CartanKilling}.

It must be stressed that supertranslations no longer commute with
themselves in the enhanced version of the conformal BMS$_{3}$ algebra,
and this is also the case for the superspecial conformal transformations.
Indeed, from the corresponding commutator in \eqref{eq:AlgebraModes},
one can read that 
\begin{align}
i\left\{ \mathcal{P}_{m},\mathcal{P}_{n}\right\}  & =i\left\{ \mathcal{P}_{m}^{0},\mathcal{P}_{n}^{0}\right\} =-\frac{3}{k}\left(m-n\right)\sum_{p}{\cal M}_{m+n-p}^{0}{\cal M}_{p}^{0}\,,\\
i\left\{ \mathcal{K}_{m},\mathcal{K}_{n}\right\}  & =i\left\{ \mathcal{P}_{m}^{1},\mathcal{P}_{n}^{1}\right\} =-\frac{3}{k}\left(m-n\right)\sum_{p}{\cal M}_{m+n-p}^{1}{\cal M}_{p}^{1}\,,
\end{align}
and hence, commutativity is lost due to nonlinear contributions of
the current generators even at the classical level, being clearly
persistent in the quantum realization. This is in stark contrast with
what occurs for the (non enhanced) conformal BMS$_{3}$ algebra in
\cite{Fuentealba:2020zkf}, since commutativity holds in that case. Indeed,
BMS$_{3}$ is a subalgebra of its conformal extension; nevertheless,
it is not a subalgebra of its enhanced conformal extension due to
the nonlinear terms in the currents.

It is worth noting that the enhanced conformal BMS$_{3}$ algebra
\eqref{eq:AlgebraModes} can also be seen as an infinite-dimensional
nonlinear extension of the AdS$_{5}$ algebra with nontrivial central
charges\footnote{An infinite-dimensional linear extension of AdS$_5$ has been proposed in \cite{Compere:2019bua,Fiorucci:2020xto}.}. Thus, the obstruction to include non trivial central extensions
for semisimple algebras, supported by a classical theorem of algebraic
cohomology (see e.g. \cite{Fuks}), can be circumvented due to the
nonlinearity of the algebra.

It is also interesting to explore whether the black hole solutions
of conformal gravity in 3D \cite{Oliva:2009hz,Oliva:2009ip,Lovrekovic:2023xsj,Lovrekovic:2024yoo}
could be endowed with an additional spin-2 field in the context of
the extension of conformal gravity of Pope and Townsend \cite{Pope:1989vj} and Grigoriev et al. \cite{Grigoriev:2019xmp}.
In order to suitably explore their properties, the asymptotic behavior
discussed here should be extended along the lines of \cite{Henneaux:2013dra,Bunster:2014mua}
so as to include the chemical potentials that correspond to the enlarged
set of global charges in \eqref{eq:QExt}.

As a final remark, it is worth exploring whether the fall-off of the
gauge fields implemented by our boundary conditions could be suitably
extended to incorporate either a finite or an infinite number of conformal
higher spin fields along the lines of \cite{Grigoriev:2019xmp} and
\cite{Pope:1989vj}, respectively. It is then natural to expect that
the full extension of the BMS$_{3}$ algebra that would emerge from
such scenarios should necessarily be nonlinear in a two-folded way.
Indeed, nonlinear extensions of BMS$_{3}$ algebra are known to appear
not only for its conformal enhancement, but also from the presence
of bosonic or fermionic higher spin fields as in \cite{Afshar:2013vka,Gonzalez:2013oaa,Gary:2014ppa,Matulich:2014hea}
and \cite{Fuentealba:2015jma,Fuentealba:2015wza}, respectively.

\acknowledgments 

We thank Luis Avilés and Joaquim Gomis for interesting remarks and
discussions. I.L. was supported by the FWF grant Hertha Firnberg T
1269-N and by the FWF grant Elise Richter V 1052-N. This research
has been partially supported by ANID FONDECYT grants N° 1211226, 1220910,
1221624. O.F. and I.L. thank the organizers of the 5th Mons Workshop
on Higher Spin Gauge Theories hosted by the Service the Physique de
l'Universe, Champs et Gravitation of the Université de Mons in January
2024, where this collaboration initiated. The authors also thank the
organizers of the ESI Programme and Workshop Carrollian Physics and
Holography hosted by the Erwin Schrödinger Institute in April 2024
in Vienna, where part of this work was carried out. The work of O.F.
was partially supported by a Marina Solvay Fellowship, as well as
by an UNAP Consolida grant of the Vicerrectoría de Investigación e
Innovación of the Universidad Arturo Prat.

\end{document}